\documentclass{article}
\usepackage[utf8]{inputenc}
\usepackage[english]{babel}
\usepackage{comment}
\usepackage{authblk}
\usepackage{setspace}
\usepackage[margin=1.4in]{geometry}
\usepackage{graphicx}
\graphicspath{ {./figures/} }
\usepackage{graphicx}
\usepackage{wrapfig}
\usepackage{lscape}
\usepackage{rotating}
\usepackage{epstopdf}
\usepackage{subcaption}
\usepackage{amsmath}
\usepackage{lineno}
\usepackage{framed}
\usepackage[most]{tcolorbox}

\colorlet{shadecolor}{orange!15}
\usepackage{mdframed}
\usepackage{multirow}
\usepackage{graphicx}
\usepackage{booktabs}

\usepackage{graphicx}
\usepackage{caption}

\usepackage{CJKutf8}
\usepackage[none]{hyphenat}
\usepackage{amssymb}

\usepackage{natbib}
\setcitestyle{authoryear,open={(},close={)}} 
\usepackage{xurl}       
\usepackage{etoolbox}
\newbool{MyRefNumbers}
\booltrue{MyRefNumbers} 

\providecommand{\keywords}[1]
{
  \small	
  \textbf{\textit{Keywords---}} #1
}

\usepackage[english]{babel}

\usepackage{amsmath}
\usepackage{graphicx}
\usepackage[colorlinks,allcolors=blue]{hyperref}
\usepackage{orcidlink} 

\title{Mastering an Accurate and Generalizable Simulation-Based Method to Obtain Bias-corrected Point Estimates and Sampling Variance for Any Effect Sizes}
\author[1,2*]{Shinichi Nakagawa\,\orcidlink{0000-0002-7765-5182}}
\author[1]{Ayumi Mizuno\,\orcidlink{0000-0003-0822-5637}}
\author[2,3]{Coralie Williams\,\orcidlink{0000-0003-1312-4953}}
\author[1]{Santiago Ortega \orcidlink{0000-0002-3518-276X}} 
\author[2,4]{Szymon M. Drobniak\,\orcidlink{0000-0001-8101-6247}}
\author[1,2]{Malgorzata Lagisz\,\orcidlink{0000-0002-3993-6127}}
\author[2,5]{Yefeng Yang\,\orcidlink{0000-0002-8610-4016}}
\author[6]{Alistair M. Senior\,\orcidlink{0000-0001-9805-7280}}
\author[7]{Daniel W. A. Noble\,\orcidlink{0000-0001-9460-8743}}
\author[1*]{Erick Lundgren \orcidlink{0000-0001-9893-3324}}
\affil[1]{Centre for Open Science and Synthesis in Ecology and Evolution (COSSEE), Department of Biological Sciences, Faculty of Science, University of Alberta, Edmonton, Canada}
\affil[2]{Evolution and Ecology Research Centre, School of Biological, Earth and Environmental Sciences, University of New South Wales, Sydney, New South Wales, Australia}
\affil[3]{School of Mathematics and Statistics, The University of New South Wales, Sydney, Australia}
\affil[4]{Institute of Environmental Sciences, Faculty of Biology, Jagiellonian University, Kraków, Poland}
\affil[4]{College of Biosystems Engineering and Food Science, Zhejiang University, Hangzhou, Zhejiang, China}
\affil[6]{Charles Perkins Centre, School of Life and Environmental Sciences, The University of Sydney, Sydney, New South Wales, Australia}
\affil[7]{Division of Ecology and Evolution, Research School of Biology, The Australian National University, Canberra, Australia}
\affil[*]{snakagaw@ualberta.ca / ejlundgr@ualberta.ca}
\date{}

\begin{document}
\maketitle
\textbf{Type of article}: Tutorial \\
\textbf{Word count}: 250 (abstract) and 6928 (main text)\\
\textbf{Number of references}: 43\\

\clearpage

\begin{abstract} 
\noindent
Meta-analyses require an effect-size estimate and its corresponding sampling variance from primary studies. In some cases, estimators for the sampling variance of a given effect size statistic may not exist, necessitating the derivation of a new formula for sampling variance. Traditionally, sampling variance formulas are obtained via hand‐derived Taylor expansions (the ``delta method''), though this procedure can be challenging for non-statisticians. Building on the idea of ``single‐fit'' parametric resampling, we introduce SAFE bootstrap: a \textbf{S}ingle‐fit, \textbf{A}ccurate, \textbf{F}ast, and \textbf{E}asy simulation recipe that replaces potentially complex algebra with four intuitive steps: fit, draw, transform, and summarise. In a unified framework, the SAFE bootstrap yields bias-corrected point estimates and standard errors for any effect size statistic, regardless of whether the outcome is continuous or discrete. SAFE bootstrapping works by drawing once from a simple sampling model (normal, binomial, etc.), converting each replicate into any effect size of interest and then calculating the bias and sampling variance from simulated data. We demonstrate how to implement the SAFE bootstrap for a simple example first, and then for common effect sizes, such as the standardised mean difference and log odds ratio, as well as for less common effect sizes. With some additional coding, SAFE can also handle zero values and small sample sizes. Our tutorial, with \texttt{R} code supplements, should not only enhance understanding of sampling variance for effect sizes, but also serve as an introduction to the power of simulation-based methods for deriving any effect size with bias correction and its associated sampling variance. \\  

\end{abstract}


\newpage
\begin{mdframed}
\section*{Highlights}

\subsubsection*{What is already known}
\begin{itemize}
    \item Sampling variances and small-sample bias corrections for effect sizes are usually obtained via delta-method expansions, which might be hard to derive for non-statisticians.
    \item Parametric bootstrapping and “single-fit” resampling have been proposed for inference, but their use for obtaining bias-corrected effect-size estimates and SEs in meta-analysis remains limited
\end{itemize}


\subsubsection*{What is new}
\begin{itemize}
    \item We formalise four steps (fit, draw, transform \& summarise) for the SAFE bootstrapping procedure, \textbf{S}ingle-fit, \textbf{A}ccurate, \textbf{F}ast, \textbf{E}asy, that yields bias-corrected point estimates and standard errors for virtually any effect size.
    \item We extend single-fit bootstrapping beyond multivariate normal outcomes to binomial and multinomial settings, incorporate paired designs via covariance structure along with a validating simulation study.
\end{itemize}

\subsubsection*{Potential impact for \textit{Research Synthesis Methods} readers}
\begin{itemize}
    \item We lower the technical barrier to introduce or adopt novel effect sizes and enables routine bias correction without new algebraic formulas.
    \item  We provide a reproducible workflow (with code and online supplement) that integrates with meta-analytic models (e.g., via the R package, \texttt{metafor}), facilitating the uptake of the SAFE bootstrapping.
\end{itemize}

\bigskip
\end{mdframed}

\clearpage
\section{Introduction}

More than four decades ago, statisticians realised that, instead of deriving sampling distributions with pencil-and-paper algebra, they could let the computer do the heavy lifting. That is, with a computer, one can resample the data many times, recalculate the statistic of interest, and use the resulting distribution for inference. The appeal of this approach, known as ``bootstrapping'', is evident as it often enables anyone to obtain standard errors and confidence intervals that remain accurate even when sample sizes are small or the underlying assumptions of deterministic estimators (e.g., normality) are not met. This simple yet powerful idea was popularised by Efron and colleagues and has since permeated many disciplines, including the field of meta-analysis \citep{efron1985bootstrap,efron1987bootstrap,tibshirani1993introduction,davison1997bootstrap}.  

Of relevance, formulas for common effect size statistics (e.g., standardised mean difference), and estimators for their sampling variance (i.e., standard errors) are frequently used when conducting meta-analysis. Formulas to estimate sampling variances are usually derived by using (first-order) Taylor expansions, also known as the ``delta method'' \citep{ver2012invented}. These formulas are elegant, yet they rely on statistical properties and were not designed for certain situations (e.g., small sample sizes and non-normality). More importantly, for non-statisticians, deriving a formula to estimate the sampling variance for a new effect size statistic seems complicated, if not impossible. Even if a formula for sampling variance is derived, how can it be verified as correct? Fortunately, the bootstrap can resolve all these issues, although its capabilities are not widely known among researchers who conduct meta-analyses. 

In this tutorial, we show how a parametric technique, termed \textbf{S}ingle-fit, \textbf{A}ccurate, \textbf{F}ast, and \textbf{E}asy-to-implement (\textbf{SAFE}) bootstrapping, can provide sampling variance and an effect size statistic without the need to derive an estimating formula.  The method is surprisingly simple. In a standard bootstrapping procedure, one redraws the data thousands of times and refits the same model each time. Unlike a standard bootstrap that refits the same model thousands of times, SAFE bootstraping fits a  ``model'' once: we draw many replicates from a statistical distribution that matches the data's structure (e.g., normal or binomial) with each draw constituting a replicate, transform each replica into an effect size, and summarise these sampled values. Moreover, using this bootstrap sample, it is straightforward to obtain a small-sample bias-corrected point estimate (for the types of bias corrected, see the next section). As with sampling variance formulas, formula-based bias corrections typically involve second-order Taylor expansions, which are also part of the delta method \citep{ver2012invented}, which are already used in some versions of the log response ratio point estimates \citep{lajeunesse2015bias}. 

The idea explored in this tutorial is not new, and previous papers have hinted at the power of this single-fit approach \citep{mandel2013simultaneous,fletcher2022single}. Yet, its potential for meta-analytic methodology and practice remains largely, if not almost completely, untapped. Our objectives are therefore threefold. First, because earlier work \citep{mandel2013simultaneous,fletcher2022single} only employed multivariate normal distributions, we extend SAFE to include binomial and multinomial distributions, catering to effect size statistics based on discrete outcomes (e.g., log odds ratio). Second, we focus on obtaining bias-corrected point estimates and sampling variances, where bootstrapping has typically been used to obtain confidence intervals. Third, we show how the idea of SAFE bootstrapping can be extended to any statistic without deriving ``new'' variance estimating formulas. We also cover how to handle edge cases, such as zero entries in discrete outcomes and negative values that should not exist in reality.

We begin with a simple example to make the SAFE bootstrap recipe accessible and concrete. We then walk through continuous-outcome effect sizes, namely log response ratio (ratio of means) and standardised mean differences, including matched-pairs designs. Next, we address discrete outcomes, such as odds and risk ratios. We also showcase the ease with which one can bootstrap less common metrics, such as the log coefficient-of-variation ratio, lnCVR \citep{nakagawa2015meta, senior2020revisiting} and the Hardy-Weinberg (dis-)equilibrium statistic \citep{wellek2010confidence}. Additionally, this tutorial is accompanied by an online supplement (\url{https://ejlundgren.github.io/SAFE/}), where not only is additional R code available \citep{rcoreteam2025r}, but also a simulation validation of the performance comparisons between conventional formulas and the SAFE bootstrap. Together, we hope that medical, biological, and social scientists, along with their students, can feel confident and ready to utilise this underappreciated yet broadly applicable method, which may encourage researchers to employ other simulation-based approaches more generally.

\section{The concept and a simple example}

\subsection{Four steps to SAFEty}

The SAFE bootstrap procedure is best understood as four steps (we denote ${\theta}$ as our effect size statistic of interest, or an estimand):

\begin{enumerate}
\item \textbf{Fit:}  
  Choose a plausible sampling model for the observed data or summary statistics (e.g.\ a normal/Gaussian model for means or a Binomial model for counts) and estimate its parameters from the observed data. This step occurs only once and is therefore considered a ``single-fit'' process. Importantly, the SAFE procedure uses the distribution of parameters or summary statistics (i.e., means and standard deviations), rather than the raw data, as is used by conventional bootstraping.

\item \textbf{Draw:}  
  Generate a large batch of simulated data directly from the fitted model in one command (e.g.\ \texttt{rnorm()} or \texttt{rbinom()}). At this stage, each draw can be considered a credible representation of ``what we might have seen'' if the study were repeated.

\item \textbf{Transform :}  
  Apply the effect-size formula to every simulated data set. Because transformations can be vectorised in software such as \texttt{R}, we turn tens of thousands of draws into tens of thousands of effect sizes very quickly.

\item \textbf{Summarise:}  
  From the resulting vector of effect size estimates $\{\theta^\ast\}$ (or $\boldsymbol{\theta}^* = \theta^*_b,\ \text{for } b = 1, \dots, B$; the vector $\{\theta^*\}$ sometimes referred to as a bootstrap cloud), we first quantify the standard deviation of the vector of effect size estimates, which is an estimate of the standard error for the parameter, denoted as $\mathrm{SE}_{\text{SAFE}}$:

\begin{equation}
    \mathrm{SE}_{\text{SAFE}} \;=\; \operatorname{sd}\!\bigl(\theta^\ast\bigr).
\label{eq:SE}
\end{equation}

Alternatively, we can calculate sampling variance as:

\begin{equation}
    \mathrm{Var}_{\text{SAFE}} \;=\; \operatorname{var}\!\bigl(\theta^\ast\bigr).
\label{eq:Var}
\end{equation}

Then, we can quantify bias in the original effect size estimate ($\hat\theta$; the hat indicates it is an estimate or estimator, an attempt to estimate the true values $\theta$, an estimand; note that, for example, Cohen's $d$ and Hedges' $g$ are estimators of the standardized mean difference) by subtracting the original estimate from the mean of the vector of effect size estimates ($\overline{\theta^\ast}$), simulated from Step 2:
  
\begin{equation}
    \widehat{\mathrm{bias}} \;=\; \overline{\theta^\ast}-\hat\theta,
\label{eq:bias}
\end{equation}

Using this estimated bias (an explanation of this formula is given below), we can obtain the bias-corrected (BC) point estimate, denoted as $\widehat{\theta}_{\mathrm{BC}}$: 
\begin{equation}
    \widehat{\theta}_{\mathrm{BC}} \;=\; \hat\theta - \widehat{\mathrm{bias}}.
\label{eq:BC}
\end{equation}

By combining Eq.~\ref{eq:bias} and Eq.~\ref{eq:BC}, we obtain:

\begin{equation}
    \widehat{\theta}_{\mathrm{BC}} \;=\;
    \hat\theta - \left(\overline{\theta^\ast}-\hat\theta\right)
    \;=\;
    2\hat\theta - \overline{\theta^\ast}.
\label{eq:BC2}
\end{equation}
\end{enumerate} 

\begin{sidewaysfigure}[ht]
  \centering
  \includegraphics[width=\linewidth]{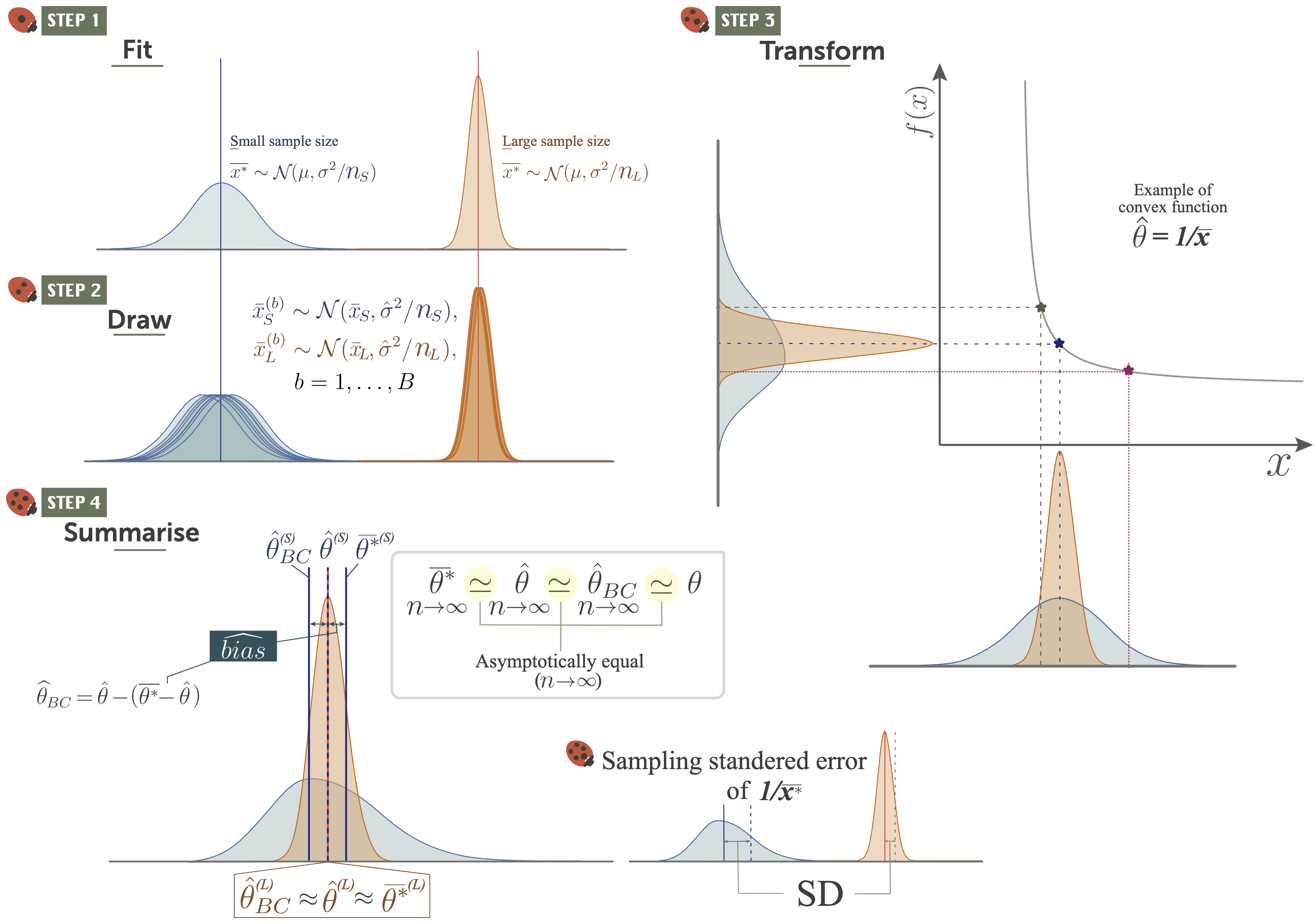}
\caption{\textbf{Visualising four steps of the SAFE bootstrapping, using an example with a convex function, $\mathbf{1 /x}$.} 
$S$ signifies a small sample size case while $L$ a large sample size case; this schematics visualise why Jensen's inequality has little impact on the large sample size case (for other symbols, see the main text). For why and how sample-sample bias correction works, see also Fig.\,\ref{fig:bias}.}
  \label{fig:safe}
\end{sidewaysfigure}

These four steps are visualised in  Fig.\,\ref{fig:safe} using a simple example presented in the next section.

This method of bias correction is elegant in its simplicity and it appears almost magical \citep[e.g.,][]{efron1987better,tibshirani1993introduction}. 
Before understanding how this bias correction works, it is essential to comprehend how such a bias arises. Most meta-analytic effect sizes are non-linear transformations of the original (raw) statistics (e.g., mean and SD).  When a transformation function $f(\cdot)$ is curved, the expected value of a transformed random variable ($x$) is not equal to the transformation of the variable's expected value (also known as Jensen’s inequality):
\begin{equation}
  E\!\bigl[f(x)\bigr] 
  \;\neq\; 
  f\!\bigl(E[x]\bigr).
\label{eq:Jensen}
\end{equation}
Previous bias corrections for the log response ratio are essentially an algebraic attempt to approximate the same offset (bias) that the SAFE bootstrap estimate also corrects \citep{lajeunesse2015bias}. By subtracting the estimated bias, we immediately obtain a second-order Taylor expansion correction that is often equivalent but without using the delta method. Importantly, with small samples, biases arising from transformation discrepancy can become non-negligible (note Fig.\,\ref{fig:safe} depicts how Jensen's inequality manifests itself and how an increase of sample size nullifies this inequality via asymptotic normality). 

\begin{sidewaysfigure}[ht]
  \centering
  \includegraphics[width=\linewidth]{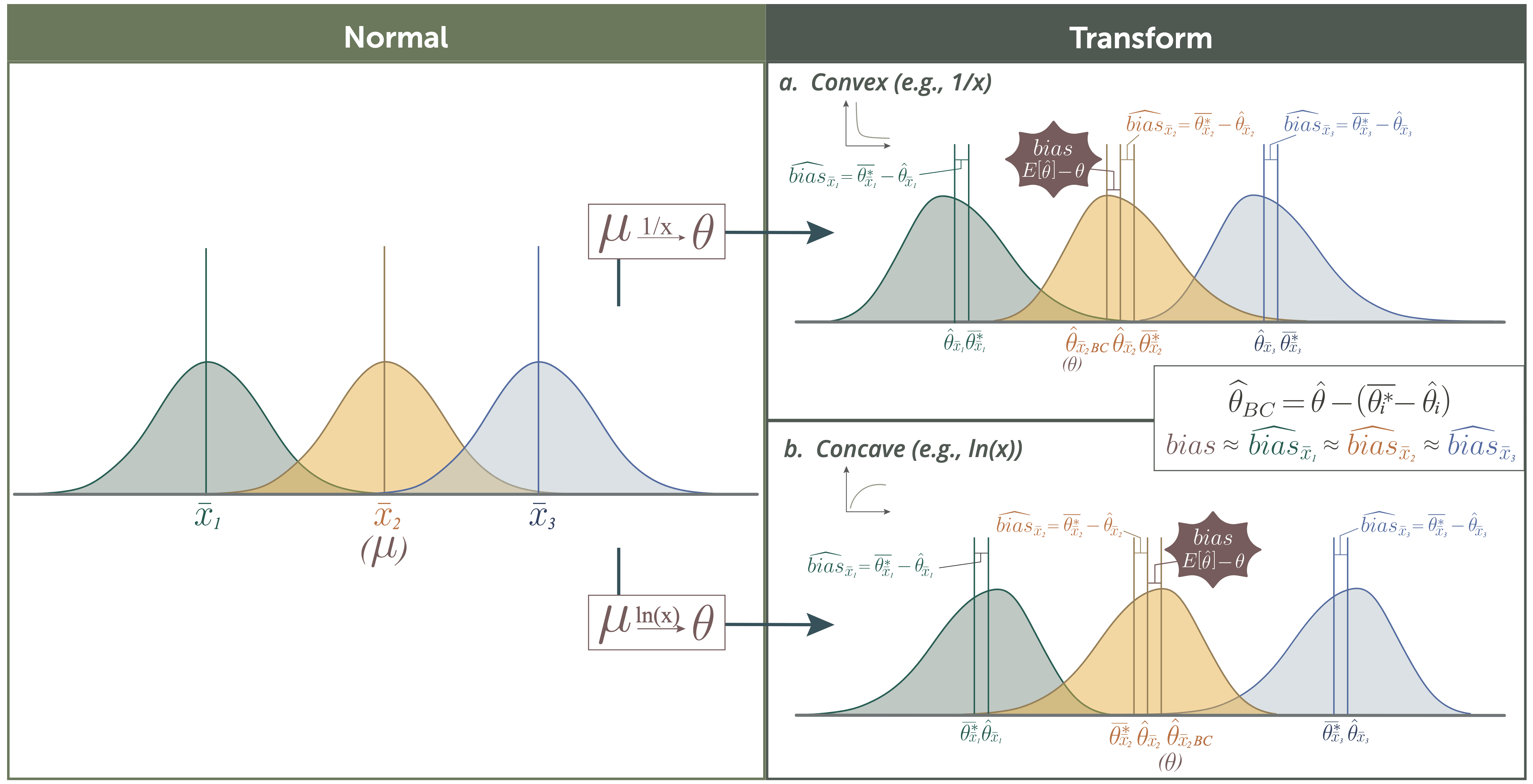}
\caption{\textbf{Bootstrap bias correction under convex vs.\ concave transforms.}
Left panel: three possible sampling distributions of the sample mean
\(\bar x_1,\bar x_2,\bar x_3 \sim \mathcal N(\mu_i,\sigma^2/n)\) with identical spread (same SE) but different locations.
Right panel: applying (a) a \emph{convex} transform (e.g., \(f(x)=1/x\)) or (b) a \emph{concave} transform (e.g., \(f(x)=\ln x\)) bends these distributions.
By Jensen’s inequality, \(E[f(\bar x)] \gtrless f(E[\bar x])\) with ``\(\ge\)'' for convex and ``\(\le\)'' for concave, yielding an upward (a) or downward (b) bias.
For each location \(i\), we simulate \(\{\bar x_i^{\ast}\}\), transform to \(\theta_i^{\ast}=f(\bar x_i^{\ast})\), and use
\(\overline{\theta_i^{\ast}} \approx E[\hat\theta_i]\) to estimate bias:
\(\widehat{\mathrm{bias}}_i=\overline{\theta_i^{\ast}}-\hat\theta_i\).
The bias‐corrected estimator is
\(\hat\theta_{BC}=\hat\theta-\widehat{\mathrm{bias}}_i\).
Because the three \(\bar xi\) distributions have the same SE, the bias (the distance between \(\overline{\theta_i^{\ast}}\) and \(\hat\theta_i\)) is essentially the same across \(i\), even though the \(\bar x_i\) (and thus \(\hat\theta_i=f(\bar x_i)\)) differ in location.
In short, the bootstrap correction is driven by curvature and uncertainty, not by where \(\bar x\) happens to sit.}
  \label{fig:bias}
\end{sidewaysfigure}

In the SAFE procedure, we directly estimate bias by drawing many potential estimates $\{\theta^{\ast}\}$ of $\hat\theta$ from our fitted sampling model ($\theta^{\ast} = \hat{\hat{\theta}}$, i.e., an estimate of an estimate). This process approximates the true sampling distribution of $\hat\theta$, and the average of those replicates, $\overline{\theta^{\ast}}$, thus approximating the expectation $E[\hat\theta]$ (or the average of many estimating attempts of $\hat\theta$, i.e., $\overline{\hat\theta}$). The statistical definition of bias is: $\mathrm{bias} = E[\hat\theta] - \theta= \overline{\hat\theta} - \theta$ (the difference between the average of many attempts of estimating what you want to estimate from a simulation, and what you want to estimate or the true value $\theta$). Then, we could estimate bias using $\widehat{\mathrm{bias}} \;=\; \overline{\theta^\ast}-\hat\theta = \overline{\hat{\hat{\theta}}}-\hat\theta$ (compare these two formulas: $\mathrm{bias} =\overline{\hat\theta} - \theta$ and $\widehat{\mathrm{bias}} = \overline{\hat{\hat{\theta}}}-\hat\theta$; the latter is a version of the former with an extra `hat' for each term). Therefore, if we take this estimated bias from the original estimate, it constitutes a bias-corrected estimate: $\widehat{\theta}_{\mathrm{BC}} = \hat\theta - \widehat{\mathrm{bias}} = 2\hat\theta - \overline{\theta^\ast}$. We have visualised how the (small-sample) bias-correction works in  Fig.\,\ref{fig:bias}, again using a simple example presented in the next section.

Before presenting this example, we briefly note why our approach earns the acronym \textbf{SAFE}: \textbf{S}ingle–fit, \textbf{A}ccurate, \textbf{F}ast, and \textbf{E}asy(-to-implement). It is because the procedure comprises:  
(i) \emph{single-fit} with a sampling model using the data;  
(ii) \emph{accuracy} through a bias-correction procedure;  
(iii) \emph{fast} estimation as one vectorised simulation replaces thousands of re-fits; and  
(iv) an \emph{easy} method, which as we show below, can use the same four steps for any effect‐size statistic. We return to each of these components as we work through our examples. 

\subsection{A simple example}

Imagine an experiment to measure an insect's performance in solving a maze (in seconds, denoted as $x$) with five insects. The mean completion time, $\bar x$, is a measure of the latency to finish a maze, so the longer it takes, the worse the performance is. To make the readout more intuitive, we may wish to take the reciprocal of the mean, so larger values mean a faster time to completion (e.g., the reciprocal of 10 sec is 0.1 while 5 sec is 0.2). This reciprocal can then be considered as a measure of ``speed''. Meta-analytically, we may wish use this reciprocal, $R$, as our new effect size statistic: 
\begin{equation}
  R \;=\; \frac{1}{\bar x}, 
\label{eq:speed}
\end{equation}
where $\bar{x}$ is the average of the time taken for maze completion for five insects. Based on this transformation, it is not immediately apparent how to calculate the sampling variance, or standard error, for $R$ (speed). Further, because the reciprocal is a curved (convex) function, the plug-in estimate $1/\bar x$ is known to be upwardly-biased (i.e., $1/\bar x >$ mean(1/$x$)) and especially so in small samples (i.e., Jensen's inequality). Now we apply our four-step SAFE procedure to $R$ (1/$x$) (Fig.\,\ref{fig:safe}).

\noindent\textbf{Step 1: Fit once.}  
The sample mean ($\bar x$) of normally distributed observations will have the following sampling distribution:
\[
  \bar x^* \;\sim\; \mathcal{N}\!\bigl(\mu,\;\sigma^2/n\bigr),
\]
where  $\bar x^*$ indicates an instance of $\bar x$, $\mu$ is the population mean, $\sigma^2$ is the population variance and $n$ is the sample size. Using the estimates from the observed sample, we plug in $\mu=\bar x$ and $\hat\sigma=\text{sd}(x)$. Note that the standard deviation of \(\bar x^*\) is \(\hat\sigma/\sqrt{n}\), not \(\hat\sigma\); the latter being the population standard deviation and the former equivalent to the standard error (i.e., the standard deviation of a sampled statistic). As mentioned above, SAFE deals with a distribution of parameters (e.g., mean), not raw data, which tend to be inaccessible to the meta-analyst. 

\noindent\textbf{Step 2: Draw once.}  
We draw $B$ surrogate sample means from this normal approximation. Larger values of $B$ are better but computationally costly. We recommend $B=100,000$, although $B=1,000,000$ could slightly improve performance, as validated in our simulation (\url{https://ejlundgren.github.io/SAFE/}), and offers an excellent balance between accuracy and efficiency. 

\vspace{6pt}
\noindent\textbf{Step 3: Transform once.}  
Each surrogate mean is converted to $R$ (a speed) by taking its reciprocal.

\vspace{6pt}
\noindent\textbf{Step 4: Summarise two metrics.}  
We obtain bias-corrected point and standard error estimates from bootstrapped data as follows:

\[
  \text{SE}_\text{SAFE}
  \;=\;
  \operatorname{sd}\!\bigl(R^{\ast}\bigr)
  \quad
  \text{and}
  \quad
  R_{\text{BC}} 
  \;=\;
  2\hat{R} - \overline{R^{\ast}}.
\]

The entire procedure is only a few lines of \textsf{R} code; note that crucially \(\text{sd}(x)\sqrt{n}\) is used in the call to \texttt{rnorm()}, ensuring that we simulate data that follow the distribution of sample means:

\begin{lstlisting}[language=R,basicstyle=\ttfamily\small]
set.seed(1)
x    <- c(11.3, 9.7, 10.4, 12.0, 9.1)    # five maze completion times
n    <- length(x)                        # number of observations in vector x
xbar <- mean(x)                          # calculate mean of observations
Rhat <- 1 / xbar                         # mean speed

## --- SAFE bootstrap ------------------------------------------------------
mu    <- xbar                            # mean time taken
sigma <- sd(x) / sqrt(n)                 # calculate SD of the mean
B     <- 1e6                             # 1 million bootstrap draws
x_star <- rnorm(B, mu, sigma)            # generating possible xbars (mean time taken)
R_star <- 1 / x_star                     # transform to speed

bias   <- mean(R_star) - Rhat            # bootstrap bias
R_BC   <- Rhat - bias                    # calculate bias-corrected speed
se_SFB <- sd(R_star)                     # calculate SAFE standard error

round(c(Rhat = Rhat, R_BC = R_BC, SE_SAFE = se_SFB), 4)
\end{lstlisting}

\noindent
The console prints the following.

\begin{verbatim}
   Rhat    R_BC SE_SAFE 
 0.0952  0.0950  0.0048 
\end{verbatim}

Three things are notable from this simple example. With a small sample size of five observations, SAFE delivers a bias correction of roughly 2\% $0.0952/0.0950 \approx 1.002$; one can immediately see the degree to which such a correction matters. The standard error emerges automatically; no delta-method derivation of \(\mathrm{Var}(1/\bar x)\) is required. Nothing in the code is specific to the reciprocal (speed; an effect size statistic).  Replace the line \texttt{R\_star <- 1 / x\_star} with any transformation, such as a standardized mean difference or an odds ratio, and the same four steps should still produce a bias-corrected estimate and a valid SE (for a simulation validation; see our online supplement: \url{https://ejlundgren.github.io/SAFE/}); incidentally, the delta-method based variance for $R$ is $\text{var}(R) = s^2/(n\bar x^4)$.

\section{Effect sizes for continuous outcomes}

In this section, we show how the same four-step SAFE logic extends to effect size statistics for continuous outcomes that compare two groups, namely, the log response ratio (ratio of means) and the standardised mean difference. First, we consider two independent groups, and then two dependent groups (paired matched or crossover design). 


\subsection{Response ratio (ratio of means: lnRoM)}


The so-called response ratio (log response ratio; also known as the log ratio of means) was first proposed by \cite{hedges1999meta} as an alternative to the standardised mean difference, which we denote as lnRoM. We note that lnRR is often used, but for this article, we reserve lnRR for log risk ratio presented in the next section. 
Imagine two groups, Group~1 ($\bar x_1 ,\, s_1 ,\, n_1$; sample mean, standard deviation and sample size) and Group~2 ($\bar x_2 ,\, s_2 ,\, n_2$). The original log response ratio can be written as:
\begin{equation}
  \ln{\text{RoM}_{1}}
  \;=\;
  \ln\left(\frac{\bar x_{1}}{\bar x_{2}}\right).
\label{eq:RoM1}
\end{equation}

This is the definition formula for lnRoM; these definition formulas are often referred to as ``plug-in'' or ``plug-ins''. A first–order Taylor expansion gives the sampling variance as:
\begin{equation}
  \operatorname{Var}\!\bigl[\ln\text{RoM}_{1}\bigr]
  \;=\;
  \frac{s_1^{2}}{n_1\bar x_1^{2}}
  +
  \frac{s_2^{2}}{n_2\bar x_2^{2}}.
\label{eq:VRoM1}
\end{equation}

Note that we label this combination of Eq.~\ref{eq:RoM1} and Eq.~\ref{eq:VRoM1} as ``\texttt{First}'' in our \texttt{R} scripts and this tutorial, although Eq.~\ref{eq:VRoM1} was not derived using the delta method and the first-order Taylor expansion. Later \cite{lajeunesse2015bias} improved both point and variance estimates by using second–order Taylor expansions (which we label as``\texttt{Second}''), which are as follows:

\begin{equation}
  \ln{\text{RoM}_{2}}
  \;=\;
   \;=\;
   \ln\left(\frac{\bar x_{1}}{\bar x_{2}}\right)
  +
  \frac{1}{2}\left(\frac{s_1^{2}}{n_1\bar x_1^{2}}
                 -
                 \frac{s_2^{2}}{n_2\bar x_2^{2}}\right).
\label{eq:RoM2}
\end{equation}

\begin{equation}
  \operatorname{Var}\!\bigl[\ln\text{RoM}_{2}\bigr]
  \;=\;
  \frac{s_{1}^{\,2}}{n_{1}\bar x_{1}^{\,2}}
  +\frac{s_{2}^{\,2}}{n_{2}\bar x_{2}^{\,2}}
  \;+\;
  \frac12
  \left(
    \frac{s_{1}^{\,4}}{n_{1}^{2}\bar x_{1}^{\,4}}
    +
    \frac{s_{2}^{\,4}}{n_{2}^{2}\bar x_{2}^{\,4}}
  \right).
\label{eq:VRoM2}
\end{equation}

For SAFE bootstrapping, we treat the pair of sample means as a bivariate
normal-distribution: 
\[
  \begin{bmatrix}
    \bar x_1^\ast \\[4pt] \bar x_2^\ast 
  \end{bmatrix}
  \;\sim\;
  \mathcal N\!\Bigl(
    \begin{bmatrix}\bar x_1\\[4pt]\bar x_2\end{bmatrix},
    \begin{bmatrix}
      s_1^{\,2}/n_1 & 0\\[4pt]
      0             & s_2^{\,2}/n_2
    \end{bmatrix}
  \Bigr).
\]
At this stage, one could use two independent normal distributions (i.e., \texttt{rnorm()} twice). However, the multivariate normal distribution has an advantage, which we will show later. A single call to \texttt{mvrnorm()} from the \texttt{MASS} package \citep{venables2013modern} draws \(B\) pairs
\((\bar x_1^{\ast},\bar x_2^{\ast})\); we transform each into
\(\ln(\bar x_1^{\ast}/\bar x_2^{\ast})\) in one vectorised step, then
compute
\[
  \mathrm{SE}_{\text{SAFE}}
  = \operatorname{sd}\left(\ln\left(\frac{\bar x_1^{\ast}}{\bar x_2^{\ast}}\right)\right),
  \quad
\text{and}
\quad
  \mathrm{lnRoM}_{\text{BC}}
  = 2\widehat{\ln\left(\frac{\bar x_1}{\bar x_2}\right)}
  -
    \overline{\ln \left( \frac{\bar x_1^{\ast}}{\bar x_2^{\ast}}\right)}.
\]

Using a standard bivariate normal distribution means simulated values may take negative values, which cannot be ln-transformed. A useful solution may be to use a truncated multivariate normal distribution, implemented in the function \texttt{rtmvnorm()} in the \texttt{R} package, \texttt{tmvtnorm}, in which we can specify zero as the lower bound \citep{wilhelm2012tmvtnorm} (see the online supplement for an example where we use \texttt{rtmvnorm()}). Below is an example \texttt{R} code, where we compare SAFE estimates to estimates from the formulas in Eqns. ~\ref{eq:RoM1} through ~\ref{eq:VRoM2}. Estimates from the second-order delta method are generally considered more accurate than those from the first-order counterpart, and we expect the SAFE estimates to be similar to the second-order estimates.


\begin{lstlisting}[language=R,basicstyle=\ttfamily\small]
set.seed(7)
## Example data (means, SDs, sample sizes) ------------------------------
x1 <- 13.4;  s1 <- 4.6;  n1 <- 18          # group 1
x2 <- 16.1;  s2 <- 3.9;  n2 <- 17          # group 2

## 1. FIRST-order (plug-in) lnRoM & SE  ---------------------------------
theta_First <- log(x1 / x2)                               # Eq. 8
var_First   <-  s1^2/(n1 * x1^2) + s2^2/(n2 * x2^2)       # Eq. 9
se_First    <- sqrt(var_First)

## 2. SECOND-order lnRoM & SE  ----------------------------------
theta_Second <- log(x1 / x2) +
                0.5 * ( s1^2/(n1 * x1^2) - s2^2/(n2 * x2^2) )  # Eq. 10

var_Second <- var_First +
              0.5 * ( s1^4/(n1^2 * x1^4) + s2^4/(n2^2 * x2^4) ) # Eq.11
se_Second  <- sqrt(var_Second)

## 3. SAFE single-fit bootstrap (multivariate Normal) -------------------
library(MASS)
mu  <- c(m1 = x1, m2 = x2)
Sig <- diag(c(s1^2/n1, s2^2/n2))            # independent sample means
B   <- 1e6                                  # number of bootstrap draws
draw <- mvrnorm(B, mu, Sig)                 # Fit-once, draw-once
theta_star <- log(draw[,"m1"] / draw[,"m2"])# Transform-once

se_SAFE       <- sd(theta_star)             # first summary: SE
bias_SAFE     <- mean(theta_star) - theta_First
theta_SAFE_BC <- theta_First - bias_SAFE    # second summary: BC point

## 4. Print tidy comparison ---------------------------------------------
results <- rbind(
  Point = c(First  = theta_First,
            Second = theta_Second,
            SAFE   = theta_SAFE_BC),
  SE    = c(First  = se_First,
            Second = se_Second,
            SAFE   = se_SAFE)
)
round(results, 4)
\end{lstlisting}

When you run the code, the console returns something like the following, where \texttt{First} represents estimates from Eq.~\eqref{eq:RoM1} and \eqref{eq:VRoM1}, \texttt{Second} represents estimates from Eq.~\eqref{eq:RoM2} and \eqref{eq:VRoM2} and \texttt{SAFE} estimates from the SAFE bootstrapping.

\begin{verbatim}
        First  Second    SAFE
Point -0.1836 -0.1820 -0.1820
SE     0.1000  0.1001  0.1007
\end{verbatim}

As you see, the point estimates from the second column (the second-order correction formulas; Eq.~\ref{eq:RoM2} \& Eq.~\ref{eq:VRoM2}) and the third column (SAFE bootstrap) match well, although SE from SAFE is slightly larger. In contrast, the point estimate from the first column (Eq.~\ref{eq:RoM1} \& Eq.~\ref{eq:VRoM1}) deviates from the other two. Note that $\text{SE}_{\text{SAFE}}$ tends to be larger when sampling sizes get smaller (which we show in the online supplement: \url{https://ejlundgren.github.io/SAFE/}).  

\subsection{Standardised mean difference (SMD)}

The standardised mean difference (SMD) remains extremely popular for continuous outcomes in ecology, psychology, and medicine. SMD may also be often referred to as Cohen's $d$ or Hedges' $g$, both of which are estimators of SMD. Although Cohen seems to have introduced SMD first \citep{cohen1969statistical}, \cite{hedges1981distribution} was the first to formalise SMD with its sampling variance.  As with lnRoM, we have two independent groups with summary statistics: Group 1 ($\bar x_{1},\, s_{1},\, n_{1}$) and Group 2 ($\bar x_{2},\, s_{2},\, n_{2}$). Following \cite{borenstein2021introduction}, we can define Cohen's $d$ as:

\begin{equation}
  d
  \;=\;
  \frac{\bar x_{1}-\bar x_{2}}{s_{\mathrm p}},
  \qquad
  s_{\mathrm p}
  =\sqrt{\frac{(n_{1}-1)s_{1}^{2}+(n_{2}-1)s_{2}^{2}}{n_{1}+n_{2}-2}},
\label{eq:d}
\end{equation}

where the term $n_{1}+n_{2}-2$ is often referred to as the degree of freedom ($df$). The sampling variance for SMD is:

\begin{equation}
  \operatorname{Var}[d]
  \;=\;
  \frac{n_{1}+n_{2}}{n_{1}n_{2}}
  \;+\;
  \frac{d^{2}}{2\,(n_{1}+n_{2}-2)}.
\label{eq:Vd}
\end{equation}

We note that this variance formula can be traced back to \cite{hedges1981distribution}, who notably used the non-central $t$ distribution in its derivation, rather than the delta method, which is commonly used elsewhere. \cite{hedges1981distribution} recognised that for small samples the expectation of \(d\) is slightly inflated \citep[see also][]{hedges1982estimation, hedges1985statistical}.  Multiplying by the correction factor: 

\begin{equation}
  J
  \;=\;
  1 \;-\;\frac{3}{\,4\,(n_{1}+n_{2}-2)\;-\;1\,},
\label{eq:HedgesJ}
\end{equation}

can alleviate this inflation; the resulting so-called Hedges' $g$ can be written as:

\begin{equation}
 g = J\left(
  \frac{\bar x_{1}-\bar x_{2}}{s_{\mathrm p}}
  \right),
\label{eq:g}
\end{equation}

And, therefore, the sampling variance of $g$ is:
\begin{equation}
  \text{Var}[g]
  \;=\;
  J^{2}\,\text{Var}[d]
  \;=\;
  J^{2}\biggl(
    \frac{n_{1}+n_{2}}{n_{1}\,n_{2}}
    +
    \frac{d^{2}}{2\,\bigl(n_{1}+n_{2}-2\bigr)}
  \biggr).
\label{eq:Vg}
\end{equation}


Given SMD includes the sample standard deviation ($s_\text{p}$ or variance $s^2_\text{p}$), we need to know the sampling variance of the variance estimate, which, assuming variances are normally distributed, is \citep{anderson2003introduction}:

\begin{equation}
  \text{Var}(\sigma^2) = \frac{2\sigma^{4}}{\,n-1\,}
\label{eq:VV}
\end{equation}

We note that variance is bounded at zero, and so when the variance is close to zero, the assumption of normality is impractical. For such cases, we discuss an alternate estimator below.

We also note that the SAFE bootstrap of SMD is by far the most challenging among those presented in this article. Now, for the SAFE procedure, we sample the quartet
\(\bigl(\bar X_{1},\bar X_{2},S^{2}_{1},S^{2}_{2}\bigr)\) jointly:

\[
\begin{bmatrix}
  \bar x_{1}^*\\[4pt]
  \bar x_{2}^*\\[4pt]
  s^{2*}_{1}\\[4pt]
  s^{2*}_{2}
\end{bmatrix}
\;\sim\;
\mathcal N \Bigl(
  {\begin{bmatrix}
    \bar x_{1}\\[4pt]
    \bar x_{2}\\[4pt]
    s_{1}^{2}\\[4pt]
    s_{2}^{2}
  \end{bmatrix}},
  \underbrace{\begin{bmatrix}
      \dfrac{s_{1}^{2}}{n_{1}} & 0                       & 0                               & 0\\[10pt]
      0                       & \dfrac{s_{2}^{2}}{n_{2}} & 0                               & 0\\[10pt]
      0                       & 0                       & \dfrac{2s_{1}^{4}}{\,n_{1}-1\,} & 0\\[10pt]
      0                       & 0                       & 0                               & \dfrac{2s_{2}^{4}}{\,n_{2}-1\,}
  \end{bmatrix}}_{\displaystyle\Sigma}
\Bigr).
\]

Off-diagonal elements of $\Sigma$ (covariance matrix) are zero because the two groups are independent, and it should be noted that the mean and variance of a normal distribution are independent of each other by definition. As before, using \texttt{mvrnorm()}, we draw $B$ independent replicates of the quartet:
($\bar x_{1}^*,\;\bar x_{2}^*,\;s_{1}^{2*},\;s_{2}^{2*}$),
Then, we vectorise the usual pooled-variance formula and compute $d$ using Eq.~\ref{eq:d}. We could have used $\chi^2$ distributions, or their multivariate version (the Wishart distribution), or \texttt{rtmvnorm()}, to sample $S_1$ and $S_2$, without zero or negative variances. 

From this vector of $d^*$ values, we read off:

\[
\text{SE}_{\text{SAFE}}
\;=\;
\operatorname{sd}\left(d^*\right)
\quad
\text{and}
\quad
d_{\mathrm{BC}}
\;=\;
2\hat{d}
\;-\;
\overline{d^*}.
\]

Below is an \texttt{R} example where we implement formulas for $d$ and $g$ along with the SAFE procedure.

\lstset{
    literate=
        {->}{$\rightarrow$}{1}
        {×}{{$\times$}}1
        {‐}{{-}}1
}

\begin{lstlisting}[language=R,basicstyle=\ttfamily\small]
set.seed(11)
## 0. Example data --------------------------------------------------------
x1 <- 13.4; s1 <- 4.6; n1 <- 18 # group 1
x2 <- 16.1; s2 <- 3.9; n2 <- 17 # group 2
df <- n1 + n2 - 2               # degrees of freedom requred for sp2

## 1. Plug‐in Cohen d & delta‐method SE -----------------------------------
sp2    <- ((n1-1)*s1^2 + (n2-1)*s2^2) / df
d_hat  <- (x1 - x2) / sqrt(sp2)            # Eq. 12
Var_d  <- (n1+n2)/(n1*n2) + d_hat^2/(2*df) # Eq. 13
se_delta <- sqrt(Var_d)

## 2. Hedges' g & its SE --------------------------------------------------
J      <- 1 - 3/(4*df - 1)
g_hat  <- J * d_hat                        # Eq. 15
Var_g  <- J^2 * Var_d                      # Eq. 16
se_g   <- sqrt(Var_g)

## 3. SAFE bootstrap on d (4‐variate) -------------------------------------
library(MASS)
mu    <- c(X1=x1, X2=x2, V1=s1^2, V2=s2^2)
Sigma <- matrix(0, 4, 4)
diag(Sigma) <- c(s1^2/n1, s2^2/n2,
                 2*s1^4/(n1-1), 2*s2^4/(n2-1))
B      <- 1e6    # the number of bootstrap
draw   <- mvrnorm(B, mu, Sigma)

# reject negative variances
draw <- draw[draw[,"V1"]>0 & draw[,"V2"]>0, ]

sp2_star <- ((n1-1)*draw[,"V1"] + (n2-1)*draw[,"V2"]) / df
d_star   <- (draw[,"X1"] - draw[,"X2"]) / sqrt(sp2_star)

se_SAFE  <- sd(d_star)
bias     <- mean(d_star) - d_hat
d_BC     <- d_hat - bias

## 4. Compare -------------------------------------------------------------
res <- rbind(
  Point = c(d   = d_hat,  g   = g_hat,  SAFE = d_BC),
  SE    = c(d =se_delta,  g   = se_g,   SAFE = se_SAFE)
)
round(res, 4)
\end{lstlisting}

We get the following in the console.


\begin{verbatim}
            d       g    SAFE
Point -0.6316 -0.6171 -0.6156
SE     0.3470  0.3391  0.3613
\end{verbatim}

We observe that the bias-corrected $d$ via SAFE is smaller than $g$ in this case, but it is closer to $g$ than $d$ (note that $g$ is an unbiased estimator). More noticeably, $\text{SE}_{\text{SAFE}}$ has a larger standard error than the other estimates, yet such differences would disappear as the sample size grows larger. A full simulation comparison is given in the online supplement (\url{https://ejlundgren.github.io/SAFE/}); in that simulation we see that the use of Wishart distributions produces better point and variance estimators than an assumed normal distribution. We also find the standard error for $d$ (Eq.~\ref{eq:Vd}) seems to perform best in combination with Hedge's $g$, and notably, \texttt{escalc()} from the \texttt{metafor} package \citep{viechtbauer2010conducting} uses this combination.

\subsection{Matched (paired) designs or repeated measures}

Many experiments measure the same individuals twice: before/after a treatment, or under two conditions. These designs are known to be more powerful because the SE can be smaller than an independent design. This is because the within‐pair correlation
$r$, can reduce the sampling variance. This reduction can be seen in action in the formula for the sampling variance for lnRoM for dependent cases (denoted by the subscript `D')\citep{lajeunesse2011meta}:

\begin{equation}
  \operatorname{Var}\!\bigl[\ln\text{RoM}_{\text{D}}\bigr]
  \;=\;
  \frac{s_{1}^{\,2}}{n_{1}\,\bar x_{1}^{\,2}}
  \;+\;
  \frac{s_{2}^{\,2}}{n_{2}\,\bar x_{2}^{\,2}}
  \;-\;
  \frac{2\,r\,s_{1}\,s_{2}}{\bar x_{1}\,\bar x_{2}\,\sqrt{n_{1}\,n_{2}}}.
\label{eq:VRoMd}
\end{equation}

Note that the within-pair correlation $r$ is ideally estimated from the original study, but may not be reported; in such cases, one can assume a specific value such as $r = 0.5$ \citep[e.g.,][]{noble2017nonindependence}. The formula for the point estimate for dependent lnRoM is as above (Eq.~\ref{eq:RoM1}). Also, with a paired design, $n_1 = n_2 = n$.

To estimate the bias-adjusted point estimates and sampling variances for dependent lnRoM via SAFE, we simply need to adjust the assumed distribution of the sample statistics in the ``fit'' step to incorporate the correlation (as covariance) into the bi-variate normal distribution from which we draw our samples as:
\[
  \begin{bmatrix}
    \bar x_{1}^*\\[4pt]\bar x_{2}^*
  \end{bmatrix}
  \;\sim\;
  \mathcal N\!\Bigl(
      \begin{bmatrix}
        \bar x_{1}\\[4pt]\bar x_{2}
      \end{bmatrix},
      \;\frac{1}{n}\,
      \begin{bmatrix}
        s_{1}^{2} & r\,s_{1}s_{2}\\[4pt]
        r\,s_{1}s_{2} & s_{2}^{2}
      \end{bmatrix}
  \Bigr).
\]

From here, the remaining steps for SAFE estimation are identical to the independent case of lnRoM.

Similarly, there exist sampling variance formulas for dependent cases of the SMD \citep[see][]{becker1988synthesizing, morris2000distribution}. However, again we can use SAFE for estimation by specifying the joint sampling distribution of the two group means and their sample variances by a quartet-variate normal with covariances \citep[cf.][]{senior2020revisiting}. A slight complication is that we need to know the sampling covariance of two variances, which is defined as \citep{anderson2003introduction}:

\begin{equation}
\text{Cov}\bigl(\sigma_1^2,\,\sigma_2^2\bigr)
= \frac{2\,\rho^2\,\sigma_1^2\,\sigma_2^2}{n-1}.
\label{eq:Cov}
\end{equation}
where $\rho$ is the true value for an estimated within-pair correlation $r$.

In the fit step of SAFE, we then draw once from:

\[
  \begin{bmatrix}
    \bar x_{1}^*\\[4pt]
    \bar x_{2}^*\\[4pt]
    s_{1}^{2*}\\[4pt]
    s_{2}^{2*}
  \end{bmatrix}
  \;\sim\;
  \mathcal{N}\!\Biggl(
    \begin{bmatrix}
      \bar x_{1}\\[4pt]
      \bar x_{2}\\[4pt]
      s_{1}^{2}\\[4pt]
      s_{2}^{2}
    \end{bmatrix},
    \begin{bmatrix}
      \dfrac{s_{1}^{2}}{n}                 & \dfrac{r\,s_{1}s_{2}}{n}           & 0                                   & 0                                   \\[6pt]
      \dfrac{r\,s_{1}s_{2}}{n}             & \dfrac{s_{2}^{2}}{n}               & 0                                   & 0                                   \\[6pt]
      0                                    & 0                                  & \dfrac{2\,s_{1}^{4}}{\,n-1\,}        & \dfrac{2\,r^{2}\,s_{1}^{2}s_{2}^{2}}{\,n-1\,} \\[6pt]
      0                                    & 0                                  & \dfrac{2\,r^{2}\,s_{1}^{2}s_{2}^{2}}{\,n-1\,} & \dfrac{2\,s_{2}^{4}}{\,n-1\,}
    \end{bmatrix}
  \Biggr),
\]

Again, the rest of the SAFE procedure remains the same, with the notable exception that $n_1 = n_2 = n$. In summary, for dependent effect sizes, once the correlation $r$ is incorporated into the covariance matrix for the ``fit'' step, the four SAFE steps remain unchanged. 

\section{Effect sizes for discrete outcomes}

Having shown that \textsc{SAFE} handles continuous‐outcome effect sizes, we now move to discrete-outcome effect size statistics, namely odds ratios (OR) and risk ratios (RR). We again use the same four steps, but the sampling model consists of a pair of independent binomial distributions. We also show how to use a continuity correction (e.g.\ adding $0.5$ to every cell value) when zero counts exist. 

\subsection{Odds ratio (lnOR)}

Many studies report a \emph{binary} outcome (e.g.\ survived/died, chose patch~A/patch~B).  
For two independent groups, we obtain the  $2\times2$ table of counts.

\[
\begin{array}{c|cc|c}
          & \text{Event} & \text{No event} & \text{Total}\\\hline
\text{Group 1} & a & b & n_{1}=a+b\\
\text{Group 2} & c & d & n_{2}=c+d\\
\end{array}
\]
with sample risks $\hat p_{1}=a/n_{1}$ and $\hat p_{2}=c/n_{2}$.
Using these four cells, we can define the log odds ratio as:
\begin{equation}
  \ln{\text{OR}}
  \;=\;
  \ln\!\left(
    \frac{a\,d}{b\,c}
  \right),
\label{eq:lnOR_def}
\end{equation}
which is a very popular effect size for this type of data.

Under a first–order Taylor expansion, the sample variance for the lnOR is \citep{woolf1955estimating, agresti2002categorical}:
\begin{equation}
  \operatorname{Var[\ln{\text{OR}}]}
  \;=\;
  \frac1a+\frac1b+\frac1c+\frac1d.
\label{eq:lnOR_var}
\end{equation}
However, this formula fails whenever $a,b,c$ or $d$ equals~$0$, as we cannot divide by~$0$.  
A popular workaround is the `continuity correction' \citep{yates1934contingency}
\[
  (a,b,c,d)\;\longrightarrow\;(a+0.5,\;b+0.5,\;c+0.5,\;d+0.5),
\]
applied before both the point estimate (Eq.~\ref{eq:lnOR_def})
and the variance (Eq.~\ref{eq:lnOR_var}) are evaluated. We note some readers may also use +1, as a continuity correction, and SAFE is compatiable with either approach.

As mentioned earlier, for SAFE we model events in each group as independent draws from two binomial distributions, as:
\[
  a^*\sim\text{Binomial}(n_{1},p_{1}),
  \qquad
  c^*\sim\text{Binomial}(n_{2},p_{2}),
\]
drawing $B$ (bootstrap) replicates $(a^{\ast},c^{\ast})$ in one line of code, and transform each pair into a \emph{continuity–corrected}
bootstrap odds ratio:

\[
  \ln{\text{OR}}^{\ast}
  =
  \ln\!\Bigl[
    \frac{(a^{\ast})\,(n_{2}-c^{\ast})}
         {(n_{1}-a^{\ast})\,(c^{\ast})}
  \Bigr].
\]

The vector of $\ln{\text{OR}}^{\ast}$ gives:

\[
\mathrm{SE}_{\text{SAFE}}=\operatorname{sd}( \ln{\text{OR}}^{\ast}) \quad
\text{and}
\quad
\theta_{\text{BC}} = 2\widehat{\ln{\text{OR}}}-\overline{ \ln{\text{OR}}^{\ast}}.
\]

Here is an example \texttt{R} script with a continuity correction of 0.5 to handle zero cells in draws.

\begin{lstlisting}[language=R,basicstyle=\ttfamily\small]
set.seed(24)

## 0. Observed 2×2 table (a b | c d) -----------------------------
a <- 2 ; b <- 20            #  2 / 22 events in Group 1
c <- 10; d <- 12            # 10 / 22 events in Group 2
n1 <- a + b
n2 <- c + d

## 1. Plug-in lnOR (no +0.5 since no zero cells) -----------------
theta_plug <- log((a * d) / (b * c))    # Eq. 20

## the delta-method SE
var_delta <- 1/a + 1/b + 1/c + 1/d      # Eq. 21
se_delta  <- sqrt(var_delta)

## 2. SAFE single-fit bootstrap (Binomial) ----------------
B <- 1e6                                # bootstrap no.
a_star <- rbinom(B, n1, a / n1)         # events  in Group 1
c_star <- rbinom(B, n2, c / n2)         # events  in Group 2
b_star <- n1 - a_star                   # non-events G1
d_star <- n2 - c_star                   # non-events G2

## Apply +0.5 only to rows with ANY zero count -------------------
needs_cc <- (a_star == 0 | b_star == 0 | c_star == 0 | d_star == 0)
a_star[needs_cc] <- a_star[needs_cc] + 0.5
b_star[needs_cc] <- b_star[needs_cc] + 0.5
c_star[needs_cc] <- c_star[needs_cc] + 0.5
d_star[needs_cc] <- d_star[needs_cc] + 0.5

theta_star <- log((a_star * d_star) / (b_star * c_star))

## 3. SAFE summaries --------------------------------------------
se_SAFE  <- sd(theta_star)
bias_hat <- mean(theta_star) - theta_plug
theta_BC <- theta_plug - bias_hat      # bias-corrected lnOR

## 4. Nicely formatted output -----------------------------------
out <- rbind(
  Point = c(First = theta_plug,
            SAFE    = theta_BC),
  SE    = c(First   = se_delta,
            SAFE    = se_SAFE)
)
print( round(out, 4) )
\end{lstlisting}

By running this script should yield the following results.

\begin{verbatim}
        First    SAFE
Point -2.1203 -1.9515
SE     0.8563  0.8714
\end{verbatim}

Both procedures provide comparable point and variance estimates, and their discrepancies disappear as sample sizes increase (see supporting simulation).

\subsection{Relative risk (risk ratio: lnRR)}

For discrete outcomes, many authors also use the relative risk, also known as the risk ratio. Using the 2 x 2 contingency table above, the log risk ratio can be defined as:
\begin{equation}
  \ln\text{RR}
  \;=\;
  \ln\!\Bigl(
    \frac{a/n_{1}}{c/n_{2}}
  \Bigr) = \ln\!\Bigl(
    \frac{p_{1}}{p_{2}}
  \Bigr)
  ,
\end{equation}

A first–order Taylor expansion for the sample variance around the observed  lnRR \citep{bailey1987confidence, agresti2002categorical}:
$ p_{1}=a/n_{1}$ and $ p_{2}=c/n_{2}$ is
\begin{equation}
  \operatorname{Var}\!\bigl[\ln\text{RR}\bigr]
  \;=\;
  \frac{1- p_{1}}{a}
  \;+\;
  \frac{1- p_{2}}{c},
  \label{eq:var_lnRR_delta}
\end{equation}
a formula that fails whenever $a$ or $c$ equals~0.  In a similar way to lnOR, yet somewhat differently, we apply the $+0.5$ continuity correction (0.5) only to $a$ and $b$ when $a = 0$, and $c$ and $d$ when $c = 0$  before evaluating Eq.~\ref{eq:var_lnRR_delta}.

For the SAFE procedure, a single call to \texttt{rbinom()} for each group suffices just as lnOR above:
\[
  a^\ast \sim \text{Binomial}(n_{1}, p_{1}),
  \qquad
  c^\ast \sim \text{Binomial}(n_{2}, p_{2}),
\]
with $b^\ast=n_{1}-a^\ast$ and $d^\ast=n_{2}-c^\ast$ filled in by definition. The four SAFE moves: fit, draw, transform, summarise, then yield:
\[
  \mathrm{SE}_{\text{SAFE}}
  \;=\;
  \operatorname{sd}\!\bigl(\ln\text{RR}^{\ast}\bigr),
  \quad
  \ln\text{RR}_{\text{BC}}
  \;=\;
  2\,\widehat{\ln\text{RR}}
  \;-\;
  \overline{\ln\text{RR}^{\ast}}.
\]

Below is an \texttt{R} example for lnRR with continuity correction of 0.5, comparing formula-based and simulation-based methods.


\vspace{6pt}
\begin{lstlisting}[language=R, basicstyle=\ttfamily\small]

set.seed(24)

## 0) Observed 2×2 table (a b | c d) ---------------------------------
a <- 2 ; b <- 20            #  2 / 22 events in Group 1
c <- 10; d <- 12            # 10 / 22 events in Group 2
n1 <- a + b
n2 <- c + d
p1 <- a / n1
p2 <- c / n2

## 1) FIRST-order (plug-in) lnRR & delta SE ---------------------------
theta_First <- log(p1 / p2)                   # Eq. 22
var_First   <- (1 - p1) / a + (1 - p2) / c    # Eq. 23
se_First    <- sqrt(var_First)

## 2) SAFE single-fit bootstrap (Binomial) ----------------------------
B <- 1e6
a_star <- rbinom(B, n1, p1)
c_star <- rbinom(B, n2, p2)

## success-only continuity correction per replicate where needed
add <- 0.5
add1 <- ifelse(a_star == 0, add, 0)
add2 <- ifelse(c_star == 0, add, 0)

p1_star <- (a_star + add1) / (n1 + 2 * add1)   # if a_star==0 -> denom n1+1
p2_star <- (c_star + add2) / (n2 + 2 * add2)   # if c_star==0 -> denom n2+1

theta_star <- log(p1_star / p2_star)
theta_star <- theta_star[is.finite(theta_star)]  # safety

se_SAFE    <- sd(theta_star)
bias_SAFE  <- mean(theta_star) - theta_First
theta_SAFE <- theta_First - bias_SAFE            # bias-corrected point

## 3) Nicely formatted comparison ------------------------------------
out_lnRR <- rbind(
  Point = c(First = theta_First,
            SAFE  = theta_SAFE),
  SE    = c(First = se_First,
            SAFE  = se_SAFE)
)
print(round(out_lnRR, 4))
\end{lstlisting}

We obtain the following. 

\begin{verbatim}
        First    SAFE
Point -1.6094 -1.4571
SE     0.7135  0.7277
\end{verbatim}

The bootstrap standard error (SE) is slightly larger, and the bias-corrected point is pulled closer to zero. Our simulation suggests that the SAFE procedure tends to give more accurate estimates. Also, it is noteworthy that for both lnOR and lnRR,  we could have used a multivariate normal to sample using modelling risks: $p_{1}$ and $p_{2}$ and their sampling variances $p_{1}(1-p_{1})/n_1$ and $p_{2}(1-p_{2})/n_2$ (for an example, see the online supplement).

\section{Less common effect sizes}

In this section, we apply the SAFE bootstrap to two lesser-known effect size statistics to demonstrate its broad capability. 

\subsection{Log coefficient‐of‐variation ratio (lnCVR)}

The coefficient of variation (CV) is defined as the ratio of the standard deviation to the mean, $\mathrm{CV}=s/\bar{x}$.  When comparing two groups in terms of variability (CV), the log CV ratio is useful, as proposed by \cite{nakagawa2015meta}:
\begin{equation}
  \ln\mathrm{CVR}_{\mathrm{1}}
  \;=\;
  \ln\!\left(\frac{s_{1}/\bar x_{1}}{\,s_{2}/\bar x_{2}\,}\right)
  \;=\;
  \ln(s_{1})-\ln(s_{2})
  \;-\;
  \bigl[\ln(\bar x_{1})-\ln(\bar x_{2})\bigr].
  \label{eq:lnCVR1}
\end{equation}

A first‐order delta–method approximation for its variance:

\begin{equation}
  \operatorname{Var}\!\bigl[\ln\mathrm{CVR}_{\mathrm{1}}\bigr]
  \;=\;
  \frac{s_1^{2}}{n_1\,\bar x_1^{2}}
  \;+\;
  \frac{s_2^{2}}{n_2\,\bar x_2^{2}}
  \;+\;
  \frac{1}{2(n_1-1)}
  \;+\;
  \frac{1}{2(n_2-1)}.
\label{eq:VlnCVR1}
\end{equation}

This accounts for both the uncertainty in the means and the sample standard deviations. 

Subsequently, \cite{senior2020revisiting} proposed second-order versions of both point and variance estimates as follows:

\begin{equation}
\ln\mathrm{CVR}_{\mathrm{2}}
=\ln\! \left( \frac{s_{1}/\bar x_{1}}{s_{2}/\bar x_{2}} \right)
\;+\;\frac{1}{2}\left(\frac{1}{n_{1}-1}-\frac{1}{n_{2}-1} \right)
\;+\;\frac{1}{2}\left(\frac{s_{2}^{2}}{n_{2}\,\bar x_{2}^{2}}-\frac{s_{1}^{2}}{n_{1}\,\bar x_{1}^{2}}\right),
  \label{eq:lnCVR2}
\end{equation}

\begin{equation}
\begin{aligned}
\operatorname{Var}\bigl[\ln\mathrm{CVR}_{\mathrm{2}}\bigr]
&=
\frac{s_{1}^{2}}{n_{1}\,\bar X_{1}^{2}}
\;+\;
\frac{s_{1}^{4}}{2\,n_{1}^{2}\,\bar X_{1}^{4}}
\;+\;
\frac{n_{1}}{2(n_{1}-1)^{2}}\\
&\quad+\;
\frac{s_{2}^{2}}{n_{2}\,\bar X_{2}^{2}}
\;+\;
\frac{s_{2}^{4}}{2\,n_{2}^{2}\,\bar X_{2}^{4}}
\;+\;
\frac{n_{2}}{2(n_{2}-1)^{2}}.
\end{aligned}
  \label{eq:VlnCVR2}
\end{equation}

They also proposed dependent cases, both the first and second order versions of sampling variances (D1 and D2, respectively), as follows:

\begin{equation}
\operatorname{Var}\bigl[\ln\mathrm{CVR}_{\mathrm{D1}}\bigr]
=
\frac{s_{1}^{2}}{n\,\bar x_{1}^{2}}
\;+\;
\frac{s_{2}^{2}}{n\,\bar x_{2}^{2}}
\;-\;
\frac{2\,r\,s_{1}\,s_{2}}{n\,\bar x_{1}\,\bar x_{2}}
\;+\;
\frac{1}{n-1}
\;-\;
\frac{r^{2}}{n-1}\,
  \label{eq:VlnCVR3}
\end{equation}

\begin{equation}
\begin{aligned}
\operatorname{Var}\bigl[\ln\mathrm{CVR}_{\mathrm{D2}}\bigr]
&=
\frac{s_{1}^{2}}{n\,\bar x_{1}^{2}}
+\frac{s_{1}^{4}}{2\,n^{2}\,\bar x_{1}^{4}}
+\frac{s_{2}^{2}}{n\,\bar x_{2}^{2}}
+\frac{s_{2}^{4}}{2\,n^{2}\,\bar x_{2}^{4}}
-\frac{2\,r\,s_{1}s_{2}}{n\,\bar x_{1}\,\bar x_{2}}\\
&\quad+\;\frac{r^{2}\,s_{1}^{2}\,s_{2}^{2}\,(\bar x_{1}^{4}+\bar x_{2}^{4})}{2\,n^{2}\,\bar x_{1}^{4}\,\bar x_{2}^{4}}
+\frac{n}{(n-1)^{2}}
-\frac{r^{2}}{n-1}
+\frac{r^{4}\,(s_{1}^{8}+s_{2}^{8})}{2\,(n-1)^{2}\,s_{1}^{4}\,s_{2}^{4}}.
\end{aligned}
  \label{eq:VlnCVR4}
\end{equation}

SAFE does not require these complex formulas; we can use the quartet with the covariances between the two means and between the two variances, where $r$ is 0 for independent groups and $r > 0$ for dependent cases (paired or matched-case designs):
\[
\begin{bmatrix}
  \bar x_{1}^*\\[4pt]
  \bar x_{2}^*\\[4pt]
  s_{1}^{2*}\\[4pt]
  s_{2}^{2*}
\end{bmatrix}
\;\sim\;
\mathcal{N}\!\Biggl(
    \begin{bmatrix}
      \bar x_{1}\\[4pt]
      \bar x_{2}\\[4pt]
      s_{1}^{2}\\[4pt]
      s_{2}^{2}
    \end{bmatrix},
  \;\;
  \underbrace{
    \begin{bmatrix}
      \dfrac{s_{1}^{2}}{n_{1}}                       & \dfrac{r\,s_{1}s_{2}}{\sqrt{n_{1}n_{2}}} & 0                                                 & 0                                                \\[6pt]
      \dfrac{r\,s_{1}s_{2}}{\sqrt{n_{1}n_{2}}}        & \dfrac{s_{2}^{2}}{n_{2}}                  & 0                                                 & 0                                                \\[6pt]
      0                                              & 0                                        & \dfrac{2\,s_{1}^{4}}{\,n_{1}-1\,}                  & \dfrac{2\,r^{2}\,s_{1}^{2}s_{2}^{2}}{\sqrt{(\,n_{1}-1\,)(\,n_{2}-1\,)}} \\[6pt]
      0                                              & 0                                        & \dfrac{2\,r^{2}\,s_{1}^{2}s_{2}^{2}}{\sqrt{(\,n_{1}-1\,)(\,n_{2}-1\,)}} & \dfrac{2\,s_{2}^{4}}{\,n_{2}-1\,}
    \end{bmatrix}
  }_{\Sigma}
\Biggr).
\]

For each of $B$ draws, we compute
\[
  \ln\mathrm{CVR}^{\ast}
  =\ln\!(\sqrt{s_{1}^{2*}}/\bar x_{1}^{\ast})-\ln\!(\sqrt{s_{2}^{2*}}/\bar x_{2}^{\ast}).
\]
The SAFE summaries are:
\[
  \mathrm{SE}_{\text{SAFE}}
  = \operatorname{sd}\!\bigl(\ln\mathrm{CVR}^\ast\bigr),
  \quad
  \ln\mathrm{CVR}_{\text{BC}}
  = 2\,\ln\mathrm{CVR}_{\text{plug-in}}
  - \overline{\ln\mathrm{CVR}^\ast}\,.
\]

Below are two examples, one assuming $r = 0$ (two independent groups) and the other $r = 0.5$ (two paired groups with a correlation of 0.5). In this instance, we create an R function that handles the independent and dependent scenarios more flexibly. 

\vspace{6pt}
\begin{lstlisting}[language=R, basicstyle=\ttfamily\small]
set.seed(1234)
library(MASS)

safe_lncvr <- function(x1, s1, n1, x2, s2, n2, r = 0, B = 1e6) {
  # --- Point estimates --------------------------------------------------
  lnCVR1 <- log(s1 / x1) - log(s2 / x2)            # Eq. 24
  lnCVR2 <- lnCVR1 +
    0.5 * (1/(n1 - 1) - 1/(n2 - 1)) +
    0.5 * (s2^2/(n2 * x2^2) - s1^2/(n1 * x1^2))    # Eq. 26
  
  # --- Variance formulas (incorporating r) ------------------------------
  if (r == 0) {
    # independent groups
    var1 <- s1^2/(n1 * x1^2) + s2^2/(n2 * x2^2) +
            1/(2*(n1 - 1))      + 1/(2*(n2 - 1))   # Eq. 25
    var2 <- s1^2/(n1 * x1^2) + s1^4/(2 * n1^2 * x1^4) + n1/(2*(n1 - 1)^2) +
            s2^2/(n2 * x2^2) + s2^4/(2 * n2^2 * x2^4) + n2/(2*(n2 - 1)^2)
                                                   # Eq. 27
  } else {
    # dependent (paired) case, assume n1 = n2 = n
    n <- n1
    var1 <- s1^2/(n * x1^2) + s2^2/(n * x2^2) -
            2*r*s1*s2/(n * x1 * x2) +
            1/(n - 1) - r^2/(n - 1)                # Eq. 28
    var2 <- s1^2/(n * x1^2) + s1^4/(2 * n^2 * x1^4) +
            s2^2/(n * x2^2) + s2^4/(2 * n^2 * x2^4) -
            2*r*s1*s2/(n * x1 * x2) +
            r^2 * s1^2 * s2^2 * (x1^4 + x2^4) / (2 * n^2 * x1^4 * x2^4) +
            n/(n - 1)^2 - r^2/(n - 1) +
            r^4 * (s1^8 + s2^8) / (2 * (n - 1)^2 * s1^4 * s2^4)
                                                   # Eq. 29
  }
  se1 <- sqrt(var1)
  se2 <- sqrt(var2)
  
  # --- SAFE bootstrap for bias correction ------------------------------
  mu <- c(x1, x2, s1^2, s2^2)
  Sigma <- matrix(0, 4, 4)
  Sigma[1,1] <- s1^2 / n1
  Sigma[2,2] <- s2^2 / n2
  Sigma[1,2] <- Sigma[2,1] <- r * s1 * s2 / sqrt(n1 * n2)
  Sigma[3,3] <- 2 * s1^4 / (n1 - 1)
  Sigma[4,4] <- 2 * s2^4 / (n2 - 1)
  Sigma[3,4] <- Sigma[4,3] <- 2 * r^2 * s1^2 * s2^2 / sqrt((n1 - 1)*(n2 - 1))
  
  draws <- MASS::mvrnorm(B, mu, Sigma)
  draws <- draws[draws[,3] > 0 & draws[,4] > 0, ]
  X1s <- draws[,1]; X2s <- draws[,2]
  S1s <- sqrt(draws[,3]); S2s <- sqrt(draws[,4])
  theta_star <- log(S1s / X1s) - log(S2s / X2s)
  
  se_safe <- sd(theta_star)
  bias    <- mean(theta_star) - lnCVR1
  lnCVR_bc <- lnCVR1 - bias
  
  c(First     = lnCVR1,
    Second    = lnCVR2,
    SAFE_BC   = lnCVR_bc,
    SE_First  = se1,
    SE_Second = se2,
    SE_SAFE   = se_safe)
}

# --- Examples -------------------------------------------------------

# Independent groups (r = 0) n can be different between two groups
res_ind <- safe_lncvr(x1 = 17, s1 = 2, n1 = 23,
                      x2 =  12, s2 = 3, n2 = 27,
                      r  = 0)

# Paired design (r = 0.5; n1 = n2 = 25) n needs to be the same
res_dep <- safe_lncvr(x1 = 15, s1 = 2, n1 = 25,
                      x2 = 10, s2 = 2, n2 = 25,
                      r  = 0.5)

# Organise into tables -----------------------------------------------
tab_ind <- rbind(
  Point = res_ind[c("First","Second","SAFE")],
  SE    = res_ind[c("SE_First","SE_Second","SE_SAFE")]
)
tab_dep <- rbind(
  Point = res_dep[c("First","Second","SAFE")],
  SE    = res_dep[c("SE_First","SE_Second","SE_SAFE")]
)
\end{lstlisting}

We should get the following (first, the independent case, and then the dependent case):

\begin{verbatim}
> round(tab_ind, 4)
        First  Second    SAFE
Point -0.7538 -0.7494 -0.7479
SE     0.2118  0.2160  0.2453
> round(tab_dep, 4)
        First  Second    SAFE
Point -0.4055 -0.4050 -0.4052
SE     0.1803  0.1853  0.2112
\end{verbatim}

This appears to be the most complex script so far, but the principles and SAFE steps remain the same. As above, the second-order delta method gives similar point estimates to SAFE. 

\subsection{Hardy–Weinberg disequilibrium} 

In population genetics, Hardy–Weinberg equilibrium (HWE) at a bi-allelic locus implies that observed genotype counts
\(\,n_{AA},n_{Aa},n_{aa}\,\) in a sample of size \(n=n_{AA}+n_{Aa}+n_{aa}\) follow proportions
\( p_1=n_{AA}/n,\;p_2=n_{Aa}/n,\; p_3=n_{aa}/n\) with \(p_1+ p_2+ p_3=1\).
The allele frequency of \(A\) is
\[
  p \;=\; \frac{2\,n_{AA}+n_{Aa}}{2\,n} \;=\; p_1+\tfrac{1}{2} p_2,
\]
so the expected heterozygote proportion under HWE is \(2 p(1- p)\).

We quantify departure from HWE (Hardy–Weinberg disequilibrium, HWD) using the relative excess of heterozygosity index as a measure of HWD \citep{wellek2010confidence},
which compares the observed heterozygosity to a symmetric homozygote-based benchmark:
\begin{equation}
  \ln\omega
  \;=\;
  \ln\!\left(\frac{ p_2}{2\,\sqrt{ p_1 p_3}}\right)
  \label{eq:lnW_def}
\end{equation}
This equals \(0\) under exact HWE (since \(2\sqrt{p_1 p_3}=2p(1-p)\) at HWE) and is label-invariant with respect to \(AA\) and \(aa\).

A first–order delta–method approximation for the sampling variance of \(\ln\omega\) is
\begin{equation}
  \operatorname{Var}\!\bigl[\ln\omega\bigr]
  \;=\;
  \frac{1}{n}\left(\frac{1}{p_2}+\frac{1- p_2}{4\,p_1\, p_3}\right).
  \label{eq:Var_lnW_delta}
\end{equation}

As you can see, these equations (Eq.~\ref{eq:lnW_def}-\ref{eq:Var_lnW_delta}) are undefined when \( p_1=0,\;p_2=0,\; \& \:p_3=0\), and as lnOR and lnRR, we can add 0.5 in such cases (i.e., continuity correction).

To generate SAFE estimates of the departure from HWE, we model genotype counts as a single draw from a multinomial distribution:
\[
  (n^*_{AA},n^*_{Aa},n^*_{aa})
  \;\sim\;
  \mathrm{Multinomial}\bigl(n;\,p_1,p_2,p_3\bigr).
\]
Then, for each of \(B\) replicates, compute
\[
  p^*_{AA}=n^*_{AA}/n,\quad
  p^*_{Aa}=n^*_{Aa}/n,\quad
  p^*_{aa}=n^*_{aa}/n,
  \qquad
  \ln\omega^* = \ln\!\Bigl(\tfrac{p^*_{Aa}}{2\,p^*_{AA}\,p^*_{aa}}\Bigr).
\]
The SE for the effect size can then be found as:
\[
  \mathrm{SE}_{\mathrm{SAFE}} = \mathrm{sd}(\ln\omega^*),
  \quad
  \ln\omega_{\mathrm{BC}} = 2\ln\omega - \overline{\ln\omega^*}.
\]

Below is an example implemented in \texttt{R}:

\vspace{6pt}
\begin{lstlisting}[language=R, basicstyle=\ttfamily\small]
set.seed(123)

## --- observed data -------------------------------------------------
nAA <- 40; nAa <- 25; naa <- 50
n   <- nAA + nAa + naa

## --- continuity correction (CC) settings ---------------------------
add <- 0.5     

## --- apply CC to observed counts only if any zero ------------------
add_obs   <- if (any(c(nAA, nAa, naa) == 0)) add else 0
n_eff_obs <- n + 3 * add_obs     # no. of genotypes = 3
p1_hat    <- (nAA + add_obs) / n_eff_obs
p2_hat    <- (nAa + add_obs) / n_eff_obs
p3_hat    <- (naa + add_obs) / n_eff_obs

## --- plug-in point & delta-method SE (use guarded p-hats) ----------
omega_hat <- log(p2_hat / (2 * sqrt(p1_hat * p3_hat))) # Eq. 30
t2        <- (1 - p2_hat) / (4 * p1_hat * p3_hat) + 1 / p2_hat
se_delta  <- sqrt(t2 / n)  # keep true n for delta SE  # Eq. 31

## --- SAFE bootstrap (Multinomial) ----------------------------------
B      <- 1e6
draws  <- rmultinom(B, n, c(p1_hat, p2_hat, p3_hat))  # mulitnomial

## Per-replicate CC if any simulated category is zero
needs_cc <- colSums(draws == 0) > 0
if (any(needs_cc)) {
  draws[, needs_cc] <- draws[, needs_cc] + add   # add 0.5 to all three categories
}
denom <- rep(n, B)
denom[needs_cc] <- n + k * add

pAA_star <- draws[1, ] / denom
pAa_star <- draws[2, ] / denom
paa_star <- draws[3, ] / denom

omegas <- log(pAa_star / (2 * sqrt(pAA_star * paa_star)))

## --- SAFE summaries ------------------------------------------------
se_SAFE  <- sd(omegas)
bias_hat <- mean(omegas) - omega_hat
omega_BC <- omega_hat - bias_hat

## --- summary table -------------------------------------------------
res <- rbind(
  Point = c(First = omega_hat, SAFE = omega_BC),
  SE    = c(First = se_delta,  SAFE = se_SAFE)
)
round(res, 4)
\end{lstlisting}

From this script, we get the following:

\begin{verbatim}
        First    SAFE
Point -1.2747 -1.2654
SE     0.2264  0.2319
\end{verbatim}

Our simulation shows that the SAFE procedure produces a less biased point estimate when the sample size is smaller, but SAFE can also overestimate SE when the sample size is small (see our online supplement).

\section{Applying SAFE functions}

So far, every example has focused on computing an effect size for single comparison, chiefly to make the four SAFE steps clear.  In reality, however, meta-analyses, comprise multiple pairwise comparisons (e.g., means, SD and n).  Fortunately, our procedure parallelises naturally. In R for example, we can wrap the SAFE function in \texttt{mapply()} or, \texttt{pmap()} from the \texttt{purrr} package \citep{wickham2025purrr}, which dispatches the function row-wise and binds the results in to a single data frame (parallelised functions are available for the effect sizes introduced above as part of the online supplement).

The R script below illustrates the workflow for log response ratios (ratio of means; \texttt{safe\_RoM}) using a meta-analytic dataset from \cite{curtis1998meta},  which is a part of the \texttt{metdat} package \citep{viechtbauer2025metdat}. As shown below, one instance of \texttt{mapply()} returns two named columns (\texttt{y\_SAFE} and \texttt{v\_SAFE}), respectively, containing the SAFE point estimates and variances for each pairwise set of summary statistics. Note that, for comparison, we also have point estimates and variances (\texttt{yi}, \texttt{vi}) obtained via \texttt{escalc()} from the \texttt{metafor} package \citep{viechtbauer2010conducting}. 

\begin{lstlisting}[language=R, basicstyle=\ttfamily\small]
## ---- 0.  preparations  -------------------------------------------------
library(metafor)
library(metadat)
set.seed(777)

## ---- 1.  SAFE log response ratios  -------------------------------------
# dataset and sampling data row indices
dat1 <- dat.curtis1998[sample(nrow(dat.curtis1998), 10), c(1, 15:20)] 

# functions for lnRoM
safe_RoM <- function(x1, x2, s1, s2, n1, n2, B = 1e6) {
  # plug-in estimate
  t0 <- log(x1/x2)
  # draws 
  x1_star <- rnorm(B, mean = x1, sd = s1/sqrt(n1))
  x2_star <- rnorm(B, mean = x2, sd = s2/sqrt(n2))
  # transform, drop any non-positive draws (if any) 
  ts <- log(x1_star[x1_star>0 & x2_star>0] / x2_star[x1_star>0 & x2_star>0])
  # bias, bias-corrected point estimate and variance 
  c(y_SAFE  = 2*t0 - mean(ts),
    v_SAFE  = var(ts)
    )
}

dat1 <- escalc(measure = "ROM",
               m1i = m1i, sd1i = sd1i, n1i = n1i,
               m2i = m2i, sd2i = sd2i, n2i = n2i,
               data = dat1, append = TRUE)

safe_out1 <- mapply(
  safe_RoM,
  x1 = dat1$m1, x2 = dat1$m2,           # means
  s1 = dat1$sd1, s2 = dat1$sd2,         # SDs
  n1 = dat1$n1,  n2 = dat1$n2,          # sample sizes
  SIMPLIFY = TRUE)

# round for easy comparisons, but usually not recommended 
dat1$yi_SAFE <- round(safe_out1["y_SAFE", ], 4)
dat1$vi_SAFE <- round(safe_out1["v_SAFE", ], 4)
\end{lstlisting}

If you run this script, we will get the following.
\begin{verbatim}
> dat1
    id      m1i       sd1i n1i     m2i      sd2i n2i     yi     vi yi_SAFE vi_SAFE 
24 242   2.2000   0.411100  10   1.330  0.189700  10 0.5033 0.0055  0.5041  0.0056 
1   21   6.8169   1.769982   3   3.945  1.115797   5 0.5470 0.0385  0.5503  0.0406 
90 739  16.5900   2.914900   6  13.800  1.592200   6 0.1841 0.0074  0.1856  0.0074 
60 456  38.3600   3.215000   2  26.490  1.790000   2 0.3702 0.0058  0.3708  0.0058 
62 458 186.2600  11.834000   2 180.480 11.867000   2 0.0315 0.0042  0.0314  0.0042 
83 726  16.9800   1.959600   6  10.960  1.935100   6 0.4378 0.0074  0.4362  0.0075 
12  96  72.5000  15.000000  24  60.500 14.000000  24 0.1809 0.0040  0.1807  0.0040 
10  87  71.7200  14.300000  16  59.940 14.280000  16 0.1794 0.0060  0.1788  0.0061 
76 615 374.4000 109.567300   5 298.200 96.374500   5 0.2276 0.0380  0.2258  0.0400 
25 254  23.1100   0.844300  22  14.940  0.844300  22 0.4362 0.0002  0.4362  0.0002 
\end{verbatim}

The same process can be implemented using the log risk ratio (lnRR; \texttt{safe\_RR}), with the dataset from \citep{graves1996vaccines}, which is also part of the \texttt{metdat} package.

\begin{lstlisting}[language=R, basicstyle=\ttfamily\small]
## ---- 2.  SAFE log risk ratios  -------------------------------------
# dataset and sampling data row indices (note should be dat.graves1996)
dat2 <- dat.graves2010[sample(nrow(dat.graves2010), 10), ]

# function for lnRR ---------------------------------------------------
# note the function is slightly different from the above 
# (it takes n1 and n2 rather than b and d
safe_RR <- function(a, c, n1, n2, B   = 1e6,
                       add = 0.5) { # continuity correction
  
  ## continuity correction --------------------------------------------
  add1 <- if (a == 0) add else 0          # only if the observed cell is zero
  add2 <- if (c == 0) add else 0
  # note the +add on *both* successes & failures
  p1_hat <- (a + add1) / (n1 + add1*2)   
  p2_hat <- (c + add2) / (n2 + add2*2)
  lnRR_raw <- log(p1_hat / p2_hat)
  # draw once 
  a_star <- rbinom(B, n1, a / n1)
  c_star <- rbinom(B, n2, c / n2)
  # transform 
  add1_star <- ifelse(a_star == 0, add, 0)
  add2_star <- ifelse(c_star == 0, add, 0)
  p1_star <- (a_star + add1_star) / (n1 + add1_star*2)
  p2_star <- (c_star + add2_star) / (n2 + add2_star*2)
  lnRR_star <- log(p1_star / p2_star)
  # bias, bias-corrected point estimate and variance 
  c(y_SAFE  = 2*lnRR_raw - mean(lnRR_star),
    v_SAFE  = var(lnRR_star)
  )
}

dat2 <- escalc(measure = "RR",
               ai = ai, ci = ci,        # events in ctrl/treat
               n1i = n1i, n2i = n2i,
               data = dat2,
               add = 0.5, to = "only0", append = TRUE)

safe_out2 <- mapply(
  safe_RR,
  a  = dat2$ai, n1 = dat2$n1,           # events + totals (grp1)
  c  = dat2$ci, n2 = dat2$n2,           # events + totals (grp2)
  SIMPLIFY = TRUE)
  
# round for easy comparisons, but usually not recommended 
dat2$yi_SAFE <- round(safe_out2["y_SAFE", ], 4)
dat2$vi_SAFE <- round(safe_out2["v_SAFE", ], 4)
\end{lstlisting}

We should now get this data frame.
\begin{verbatim}
> dat2
             study ai    n1i ci   n2i      yi     vi yi_SAFE vi_SAFE 
12  Azurin 1965-ii 92 148100 52 48933 -0.5369 0.0301 -0.5410  0.0308 
16   Saroso 1978-i 18 156300 19 79250 -0.7332 0.1082 -0.7316  0.1186 
9    PCC 1973a-iii 36  44450 23 11050 -0.9439 0.0711 -0.9527  0.0756 
8     PCC 1973a-ii 33  45750 22 11050 -1.0153 0.0756 -1.0233  0.0806 
3    Mosley 1970-i 32  11491 13  3810 -0.2031 0.1078 -0.2286  0.1205 
17  Saroso 1978-ii 10 155600 18 79250 -1.2625 0.1555 -1.2367  0.1819 
1    Oseasohn 1965  8   6956 33  7103 -1.3962 0.1550 -1.3413  0.1911 
15        PCC 1968 54 268700 41 90900 -0.8084 0.0429 -0.8116  0.0443 
10    PCC 1973a-iv 26  44700 22 11050 -1.2305 0.0838 -1.2341  0.0896 
13 Azurin 1965-iii 71 143600 52 48933 -0.7651 0.0333 -0.7679  0.0342   
\end{verbatim}

As you can see, SAFE provides very similar point estimates and variance estimates as \texttt{escalc()}, though the variance estimates are often larger (more conservative), and this discrepancy disappears as the sample size grows. In practice, for many existing studies, SAFE may not change the qualitative conclusions. However, the SAFE procedure future-proofs the analysis for new effect-size statistics. Additionally, one can compare the point estimate and variance derived from the delta method for new effect size statistics with those obtained using the SAFE bootstrap, as we have done in this tutorial. 

\section{Concluding remarks}

\noindent
In this tutorial, we have introduced the Single-fit, Accurate, Fast, and Easy (SAFE) bootstrap that builds on previous single-fit parametric methods \citep{mandel2013simultaneous, fletcher2022single}. We have also demonstrated that it is a universally applicable and intuitive approach. By following a straightforward four-step recipe (fit, draw, transform, and summarise), researchers can liberate themselves from the delta-method-based (approximation) formulas \citep{ver2012invented}. Furthermore, SAFE can complement such formulas by checking when they may not work well (e.g., how small is a small sample size for a particular effect size statistic?). This approach democratises the process of obtaining the sampling variance of effect statistics, allowing any practitioner to derive a bias-corrected point estimate and standard error (variance) for virtually any conceived effect size statistic, with little more than basic coding and knowledge about statistical distributions. Although it is not the focus of our paper, our simulation validation has shown that SAFE point estimates are often superior to the plug-in and small-sample corrected formulas, especially when the sample size is small. In line with our results, the simulation work of \cite{fletcher2022single} demonstrated that confidence intervals (CIs) derived from the SAFE procedure (i.e., single-fit bootstrapping) performed better than those obtained using the delta method, recommending SAFE over the delta method for CI constructions. Yet, we also note that delta-method-based variance estimators could outperform their SAFE counterparts in some scenarios; SAFE sampling variance tends to be higher than that of the delta method (see the online supplement: \url{https://ejlundgren.github.io/SAFE/}).

Here, we have also shown that the SAFE procedure can be applied to familiar effect size statistics, such as lnRoM, SMD, lnOR, and lnRR. Yet the true potential of the SAFE bootstrap lies in its adaptability. The same reasoning and coding pipeline can be applied to less common but valuable metrics, such as lnCVR \citep{nakagawa2015meta, senior2020revisiting} or the Hardy-Weinberg equilibrium statistic \citep{wellek2010confidence}, as illustrated. Additionally, for some effect sizes, variance estimators are not derived for paired or dependent designs; however, this is not an issue for the SAFE approach, which can readily handle dependence. This flexibility allows researchers to choose the most suitable effect size for their specific question, rather than being restricted to effect sizes for which estimators of the sampling variance exist. Furthermore, we note that SAFE is useful for converting effect size statistics; correlation, lnOR, and SMD are convertible to each other \citep{borenstein2021introduction}, but these formulas are likely to introduce bias due to their non-linear nature. It is also an interesting future avenue of research to apply the SAFE procedure with the concept of ``smooth variance estimators'', where sampling variance is calculated by averaging observed parameter values. For example, this approach can be applied to CV, which is a part of the sampling variance formula for lnRoM, which improves the estimation of overall (meta-analytic) means \citep{nakagawa2023robust}. 

With the SAFE procedure, the initial ``fit'' step involves choosing an appropriate sampling model for the sample statistics, which requires careful consideration and statistical understanding. Here, we have made heavy use of the multivariate normal distribution and the binomial distribution. However, one can utilise any distribution that accurately reflects the data's nature; for example, instead of a multivariate normal distribution, one could use a multivariate log-normal distribution or a Gamma distribution. Beyond step 1, the parts of the SAFE approach are largely identical across effect size types, with a few edge cases that require attention (e.g., negative variances and zero cells; see online supplement). A notable corollary is that the distribution obtained via SAFE can be likened to a Bayesian posterior distribution, bridging the gap between frequentist and Bayesian statistics. Indeed, Bayesian inference often involves processing the posterior distribution in a manner similar to that of the SAFE bootstrapped data \citep{efron2012bayesian}. 

Ultimately, we believe the SAFE bootstrap offers a reliable and accessible tool to add to the meta-analyst's toolkit. We do not aim for SAFE to replace the elegant formulas derived by statisticians to estimate effect sizes. Rather, we see SAFE as a useful tool that can provide practitioners with a robust simulation-based procedure that is computationally efficient and easy to grasp for novel measures of effect. Furthermore, SAFE can foster a deeper and more intuitive understanding of sampling distributions and bias, thereby improving the immediate practice of calculating effect sizes \citep[e.g.,][]{noble2019plastic,joshi2022cluster,jarrett2025meta}. Therefore, the path to more accessible and generalisable meta-analytic inference is, as we have shown, a SAFE one.

\subsection*{Acknowledgment}
SN acknowledges that conversations with David Fletcher and Wolfgang Viechtbauer have inspired the idea of this manuscript, and SN would like to thank them. SN, SO, AM, and EL were supported by the Canada Excellence Research Chair Program (CERC-2022-00074). YY, ML and SN were supported by the Australian Research Council Discovery Project (DP230101248). AMS and DWAN were supported by an Australian Research Council Future Fellowship (FT230100240, FT220100276, respectively). 

\subsection*{Author Statement}
Conceptualization: All.
Formal analysis: EL, SN.
Funding Acquisition: SN, ML.
Investigation: EL, SN.
Methodology: EL, SN.
Project administration \& Supervision: SN.
Validation: EL, DWAN, AMS, SO.
Visualization: AM.
Writing - Original Draft: SN, EL.
Writing - Review \& Editing: All.

\subsection*{Conflict of Interest}
The authors declare no conflict of interest

             

\clearpage

\clearpage

\end{document}